\documentclass[11pt, a4paper]{article}

\usepackage{color}
\usepackage{graphicx}
\usepackage{amssymb}
\usepackage{amsmath}
\usepackage[english]{babel}
\usepackage[T1]{fontenc}
\usepackage[noadjust]{cite}
\usepackage{hyperref} 
\usepackage{cancel}
\usepackage[normalem]{ulem}

\usepackage[utf8]{inputenc}
\usepackage{authblk}

\newcommand\eeq{\end{equation}}
\newcommand\beq{\begin{equation}}
\newcommand\eea{\end{eqnarray}}
\newcommand\bea{\begin{eqnarray}}

\begin{document}

\linespread{1.1}

\title{ \color{red} \bf Bounds on warm dark matter from \\ Schwarzschild  primordial black holes}

\author[1]{ {\Large J\'er\'emy Auffinger} \thanks{auffinger@ipnl.in2p3.fr}}
\author[2]{ {\Large Isabella Masina} \thanks{masina@fe.infn.it}}
\author[3]{ {\Large Giorgio Orlando} \thanks{g.orlando@rug.nl}}

\affil[1]{\small Univ. Lyon, Univ. Claude Bernard Lyon 1, CNRS/IN2P3, IP2I Lyon, UMR 5822, F-69622,
Villeurbanne, France } 
\affil[2]{\small Dip. di Fisica e Scienze della Terra, Ferrara University and INFN, Via Saragat 1, 44122 Ferrara, Italy }
\affil[3]{\small Van Swinderen Institute for Particle Physics and Gravity, University of
Groningen, Nijenborgh 4, 9747 AG Groningen, The Netherlands} 

\date{}

\maketitle

\begin{abstract}
We consider light dark matter candidates originated from the evaporation of Schwarzschild primordial black holes, with masses in the range $10^{-5}-10^9$ g. These candidates are beyond Standard Model particles with negligible couplings to the other particles, so that they interact only gravitationally. Belonging to the category of warm dark matter, they nevertheless spoil structure formation, with a softer impact for increasing values of the candidate spin. Requiring such candidates to fully account for the observed dark matter, we find that the scenario of black hole domination is ruled out for all spin values up to 2. For the scenario of radiation domination, we derive upper limits on the parameter $\beta$ (the primordial black hole energy density at formation over the radiation one), which are less stringent the higher the candidate spin is.
\end{abstract}

\linespread{1.2}

\vskip 1.cm

\section{Introduction}

Dark matter (DM) candidates must be beyond Standard Model (SM) particles, neutral and stable. Having so far escaped detection, they must have tiny interactions with SM particles. It would be even possible that they interact only gravitationally. 

A possible production mechanism for DM particles, taking place in the early universe, is via evaporation \cite{Hawking:1974sw} of primordial Black Holes (BHs), with masses in the broad range $10^{-5}-10^9$ g. In this case, all particles with mass below the Hawking temperature of the BH are emitted, with weights simply given by their number of degrees of freedom (dof). 
It has been proposed that the particles produced via the evaporation mechanism might be responsible for the excess of baryons over anti-baryons \cite{Hawking:1974rv, Zeldovich:1976vw}, for the observed dark matter abundance \cite{Fujita:2014hha, Lennon:2017tqq, Morrison:2018xla}
and, if sufficiently light, also for dark radiation \cite{Lennon:2017tqq, Hooper:2019gtx, Lunardini:2019zob, Hooper:2020evu}. 
Apart from the case of gravitino production \cite{Khlopov:2004tn, Khlopov:2008qy}, the primordial BH density at formation for the range $10^{-5}-10^9$ g is at present unconstrained, as reviewed \emph{e.g.} in ref.\,\cite{Carr:2020gox}. However, ref.\,\cite{Papanikolaou2020} (see also ref.\,\cite{Domenech:2020ssp}) derives an upper bound on the fraction of the universe collapsed into primordial BHs in this very mass range from possible backreaction gravitational waves from primordial BHs. Ref.\,\cite{Dai:2009hx} considers constraints on DM particles charged under a hidden gauge group.  

Depending on the fraction of primordial BHs at formation with respect to radiation $\beta$, there is a possibility that the universe was BH dominated before the evanescence of the BHs \cite{Barrow:1990he, Baumann:2007yr, Fujita:2014hha}: this situation is referred to as BH domination. The case in which the BHs evaporate before they dominate the energy content of the universe is called radiation domination.

Fujita \emph{et al.} \cite{Fujita:2014hha}
calculated the contribution to DM by primordial BH evaporation into new particles beyond the SM: they found that a significant contribution to DM could come from stable particles that are either superheavy or light, that is with masses in the MeV range. In the light case, DM candidates would be warm, while in the superheavy case they would be cold. Exploiting the warm DM velocity constraints available at that time \cite{Viel:2005qj}, ref.\,\cite{Fujita:2014hha} first discussed also the lower limits on the mass of the light DM candidates, using an order-of-magnitude argument essentially based on the geometrical optics approximation for Hawking radiation. This approximation ignores the low energy suppression in the greybody factors \cite{Starobinsky:1974, Page:1976df}, accounting quite well for the case in which the warm DM candidate has $s=0$, but missing to reproduce the case of different spins. For an up-to-date presentation of this argument see \cite{Masina:2020xhk, Baldes:2020nuv}.

A more sophisticated analysis was done by Lennon \emph{et al.} \cite{Lennon:2017tqq}. They also adopted the geometrical optics approximation, but included the redshift effect in the calculation of the momentum distribution of the emitted particles. Their result is an estimate of the number of particles that are still relativistic, with a spin dependence reintroduced \emph{a posteriori} and based on greybody factors
derived from the older literature \cite{Page:1976df, MacGibbon:1990zk}. As a rough-and-ready criterion for successful structure formation, they impose that when the temperature of the universe drops below $1$ keV (at which stage the horizon mass is about $10^9$ solar masses), less than $10\%$ of the DM is relativistic. The result of this ingenious, but quite arbitrary, argument is that, for BH domination, warm DM candidates with $s \leq 1$ are excluded, those with $s=3/2$ are marginally allowed, while those with $s=2$ naively survive. Summarizing, for the lower spin values (say $s=0,1/2,1$), the order of magnitude results of ref.\,\cite{Fujita:2014hha} were confirmed by ref.\,\cite{Lennon:2017tqq}, but the latter analysis was however not fully conclusive for the higher spins ($s=3/2,2$).

The more recent analysis of Baldes \emph{et al.} \cite{Baldes:2020nuv} goes some step further. As suggested in \cite{Lennon:2017tqq}, they include the redshift effect in the momentum distribution of the emitted particles at evaporation, 
and derive the related phase-space distribution as an input for the Boltzmann code \texttt{CLASS} \cite{Lesgourgues:2011re,Blas:2011rf,Lesgourgues:2011rh}. The latter allows to extract the matter power spectrum for warm DM from primordial BHs, and to compare it to the standard cold DM case thanks to the transfer function. This enables to constrain warm DM from primordial BHs using the structure formation bounds from Lyman-data already derived for the well known case 
of DM thermal relics. The analysis of ref.\,\cite{Baldes:2020nuv} however relies on the geometrical optics approximation and, in particular, provides quantitative results only for the $s=1/2$ case, which agree with previous order-of-magnitude estimates \cite{Fujita:2014hha,Masina:2020xhk}, also based on the geometrical optics approximation. The case for the higher spins could thus not be quantitatively clarified (apart from a qualitative mention of the greybody effects in appendix A of ref.\,\cite{Baldes:2020nuv}) with respect to the results of ref.\,\cite{Lennon:2017tqq}.

Given the present lack of robust results about the fate of warm DM candidates with high spin values, we think it would be useful and timely to make a dedicated study. The aim of this work is precisely to provide a complete and updated study on the viability of warm DM candidates from the evaporation of primordial BHs.

In order to numerically account for the greybody factors associated to the different spins, we use the recently developed and publicly available code \texttt{BlackHawk} \cite{Arbey:2019mbc}. We also compare the numerical results from \texttt{BlackHawk} with the analytical ones derived in the geometrical optics approximation. Taking into account the redshift effects as suggested in ref.\,\cite{Lennon:2017tqq}, we study the impact on structure formation by calculating the transfer function with \texttt{CLASS}, as suggested in ref.\,\cite{Baldes:2020nuv}.
We derive the transfer function for all spins values, finding that, assuming BH domination, the scenario of warm DM from primordial BHs is excluded for all spins and for all BH masses in the range $10^{-5}-10^9$ g.
Our results for the $s=0$ case, agree with previous order-of-magnitude estimates \cite{Fujita:2014hha,Masina:2020xhk}. 
For radiation domination, we derive the upper limits on $\beta$ (or, equivalently, on the warm DM mass) for the various warm DM spins. For the case $s=1/2$ (the only for which the comparison is possible), we find conceptual differences with respect to the results of ref.\,\cite{Baldes:2020nuv}, but substantial numerical agreement.

In this work we consider BH evaporation as the only production mechanism. The consequences of allowing for other production mechanisms have been recently explored in refs.\,\cite{Gondolo:2020uqv} and \cite{Bernal:2020ili, Bernal:2020bjf}. For a mixed model of DM production, ref.\,\cite{Gondolo:2020uqv} proved that a primordial BH dominated period of DM creation by evaporation cannot explain the abundance observed today. For an updated analysis of the possibility that the matter-antimatter asymmetry is due to particles produced by primordial BHs evaporation, 
we refer the interested reader to ref.\,\cite{Hooper:2020otu} for GUT baryogenesis, and to ref.\,\cite{Perez-Gonzalez:2020vnz} for leptogenesis. DM and baryogenesis in the case of stable remnants from thermal 2-2-holes have been studied in ref.\,\cite{Aydemir:2020pao}.

The paper is organized as follows. In sec.\,2 we introduce our notation and review basic ideas about formation and evaporation of primordial BHs. In sec.\,3 we discuss the instantaneous primary spectrum for the emitted particles. In sec.\,4 we discuss the dynamics of the primordial BH abundance. Sec.\,5 deals with the momentum distribution at evaporation, and sec.\,6 with the calculation of the DM phase space distribution. The calculation of the DM abundance  and the impact on structure formation are presented in secs.\,7 and 8 respectively. The discussion of the results and our conclusions are presented in sec.\,9.

In order to have a better control of our formulas for dimensional analysis and numerical computations, we do not use natural units.

\section{Preliminaries on primordial BHs}  

The proposal of the early existence of collapsed objects, later called primordial BHs, dates back to 1967 \cite{ZelNov:1967}. 
The formation of primordial BHs from early universe inhomogeneities was considered in refs.\,\cite{Hawking:1971ei, Carr:1974nx, Hawking:1974sw}. However, since inflation removes all pre-existing inhomogeneities, any cosmologically interesting primordial BH density has to be created after inflation. Various mechanisms have been proposed, as for instance: that they formed from large inhomogeneities arising through quantum effects during inflation or that some sort of phase transition may have enhanced primordial BHs formation from primordial inhomogeneities or triggered it. We refer to \cite{Carr:2020gox, Khlopov:2013ava} for reviews of these proposals, with proper references to the associated literature.
In the following, we present our notation for the early universe dynamics and review the primordial BH formation and evaporation mechanisms.

\subsection{Radiation dominated era }

According to the first Friedmann equation (neglecting the curvature and cosmological constant terms), 
the early universe evolution is described by
\beq
 \left( \frac{\dot a}{a} \right)^2   \equiv H(t)^2 =\frac{8 \pi G}{3 }   \rho(t)\,,
\label{eq-F1}
\eeq
where $a(t)$ is the scale factor, $H(t)$ is the Hubble parameter, $\rho(t)$ is the energy density of the universe 
and $G$ is the Newton gravitational constant, $G \simeq 6.674 \times 10^{-11} \,\rm{ m^3 / (kg\,s^2)}$. 
Instead of $G$, it might be convenient to rather use the Planck mass, $M_{ Pl} \equiv \sqrt{ \hbar c / G } \approx 1.221 \times 10^{19}\, \rm{GeV}/c^2  \approx  2.176 \times 10^{-8} $ kg. 

In the early hot and dense universe, it is appropriate to assume an equation of state corresponding to a fluid of radiation (or relativistic particles). 
During radiation domination, $\rho \propto a^{-4}$, $a(t) \propto t^{1/2}$, and
\beq
H(t)= \frac{1}{2 t }\,.
\label{eq-raddom}
\eeq

At relatively late times, non-relativistic matter eventually dominates the energy density over radiation. 
A pressureless fluid 
leads to the expected dependence $\rho \propto a^{-3}$, $a(t) \propto t^{2/3}$, and
\beq
H(t) = \frac{2}{3 t} \,.
\label{eq-matdom}
\eeq

The radiation energy density (at high temperatures) can be approximated by including only those particles which are in thermal equilibrium and have masses below the temperature $T$ of the radiation bath
\beq
 \rho_R = \frac{\pi^2 g_*(T)  } {30}  \frac{(k_B T )^4 }{(\hbar \,c)^3 \,c^2} \, , \quad
 g_*(T)= \sum_B g_B + \frac{7}{8}\sum_F g_F \, , 
 \label{eq-rhoRT}
\eeq
where $k_B$ is the Boltzmann constant, $k_B \simeq 8.617 \times 10^{-5} $ eV/K, 
$\hbar$ is the reduced Planck constant, $\hbar \simeq 6.582 \times 10^{-16}$ eV\,s, 
$c$ is the velocity of light in the vacuum, $c \simeq 2.998 \times 10^{8}$ m/s,
and $g_{B(F)}$ is the number of degrees of freedom (dof) of each boson (fermion). 

Below the electron mass, 
only the photon ($g_\gamma=2$) and three light left-handed neutrinos contribute, 
so that $g_*(T)=7.25$. Below the muon mass, 
also the electron (and the positron) has to be included, 
so that $g_*(T)=10.75$. 
For the full SM, here defined including three light left-handed neutrinos, $g_*(T)= 106.75$. 
Adding to the SM three light right-handed neutrinos (as in the case of neutrinos with Dirac nature or in the case of a low-scale seesaw mechanism),  $g_*(T)= 112$.
At higher temperatures, $g_*(T)$ will be model-dependent. 
Including the massless graviton ($g_G=2$) has the effect of adding $2$ units to the previously mentioned values of $g_*(T)$.

\subsection{Formation of primordial BHs}

As reviewed for instance in ref.\,\cite{Carr:2020gox}, if a primordial BH forms at the time $t_f$ during the radiation dominated era, typically its mass is close to the value enclosed by the particle horizon near the end of inflation
\beq
M_{BH} =  \gamma   \frac{4 \pi}{3}  \rho_R(t_f) \left( 2\, c \,t_f \right)^3 =\gamma   \frac{4 \pi}{3}  \rho_R(t_f)  \left( \frac{ c }{H(t_f)}   \right)^3\,,
\label{eq-MPBH}
\eeq
where $\gamma \lesssim 1$ is a numerical factor that depends on the details of the gravitational collapse, $\rho_R(t_f)$ and $H(t_f)$ are respectively the radiation density and the Hubble parameter at the formation of the BH, and in the last equality we used eq.\,(\ref{eq-raddom}). Using eq.\,(\ref{eq-F1}), we can also write 
\beq
 M_{BH} =   \frac{\gamma}{2} \frac{(M_{Pl}c^2) ^2}{ \hbar \,H(t_f)}\frac{1}{c^2} \approx  \gamma   \frac{10^{10} \,\rm{GeV}} {\hbar \,H(t_f)}  10^4 \,\rm{g}  \gtrsim \frac{\gamma}{3} \,{\rm  g} \, ,
\label{eq-PBH}
\eeq
where the last lower bound follows from the fact that CMB observations put an upper bound on the Hubble scale during inflation, $\hbar H_I \lesssim 3 \times 10^{14}$ GeV at $95\%$ C.L. \cite{Akrami:2018odb}, and $H(t_f) \lesssim H_I$. In the literature the value $\gamma=1/(3\sqrt{3})\approx 0.2$ is usually taken as reference value \cite{Carr:2020gox}; in this case the lower limit would become $M_{BH} \gtrsim 0.07$ g. For values of $\gamma$ smaller than $0.2$, the lower bound on the BH mass would get accordingly weaker. However, this constraint applies only to conventional inflationary scenarios, \emph{i.e.} the standard slow-roll models of inflation with Einstein gravity (see \emph{e.g.} the review \cite{Baumann:2009ds} and references therein). In more sophisticated scenarios, like \emph{e.g.} the recent model \cite{Aoki:2020ila}, the scale of inflation can not be determined by CMB observations. In any case, the mass of primordial BHs should be larger than the Planck mass, namely $M_{BH}\gtrsim 10^{-5}$\,g. As is well known, there is also an upper bound on the abundance of primordial BHs of mass $M_{BH} \gtrsim 10^9$ g (not a theoretical bound), because of their effects on BBN yields, see \emph{e.g.} ref.\,\cite{Keith:2020jww} for a recent analysis. The range of primordial BH masses between these bounds is at present generically unconstrained \cite{Carr:2020gox}.

Recalling eq.\,(\ref{eq-raddom}), the primordial BHs formation time is easily calculated from eq.\,(\ref{eq-PBH})
\beq
\frac{t_{f}}{\hbar}=\frac{1} {\gamma}  \frac{M_{BH} c^2}{ (M_{Pl} c^2)^2  } \,.
\label{eq-tfBH}
\eeq
As for the radiation temperature at formation,
combining eqs.\,(\ref{eq-F1}), (\ref{eq-rhoRT}) and (\ref{eq-PBH}), we have 
\beq
k_B T_R(t_{f})  
 = \left(  \frac{45 \gamma^2  } {16 \pi^3 g_*(t_{f}) }   \right)^{1/4}  \left( \frac{M_{Pl}  }{M_{BH}} \right)^{1/2}\, M_{Pl} c^2 \,.
 \label{eq-Tf}
\eeq
The temperature and the time at formation of primordial BHs, as a function of the their mass at formation, are plotted \emph{e.g.} in fig.\,1 of ref.\,\cite{Masina:2020xhk}.

It is useful to introduce the parameter $\beta$ defined as the BH energy density over the radiation energy density at the formation time
\beq
\beta \equiv \frac{\rho_{BH}(t_f)}{\rho_R(t_f)}= M_{BH} \frac{n_{BH}(t_f)}{\rho_R(t_f)} \,,
\label{eq-defbeta}
\eeq
where $n_{BH}(t_f)$ is the primordial BH number density at formation and the last equality holds only for a monochromatic mass distribution.

\subsection{Evaporation of primordial BHs } 

Here we review the basic formulas describing the evaporation mechanism, in the case of a Schwarzschild (that is, uncharged and non-rotating) primordial BH.

Consider a Schwarzschild BH of mass $M_{BH}(t)$ (we neglect the time dependence only when we refer to the formation time). Hawking radiation mimics thermal emission from a blackbody with a temperature $T_{BH}(t)$, 
given by \cite{Hawking:1974sw}
 \beq
k_B T_{BH}(t) 
=\frac{ 1}{8 \pi } \frac{  (M_{Pl} c^2)^2}{ M_{BH}(t) \,c^2}\,  .
\label{eq-TBH}
\eeq
Hereafter we denote by $T_{BH}$ the Hawking temperature at formation, namely $T_{BH}=T(t_f)$. As discussed \emph{e.g.} in \cite{MacGibbon:1991tj}, at the time $t$, such a hole emits particles of type $i$ and spin $s_i$ and total energy between $(E, E+dE)$ at a rate, per dof, given by
\beq
\frac{1}{g_i} \frac{d^2N_i}{dt \,dE}  =  \frac{d^2N}{dt \,dE} 
= \frac{1}{2 \pi \hbar } \, \Gamma_{s_i}(E,T_{BH}((t)) \,  \frac{1}{e^\frac{E}{k_B T_{BH}(t)} - (-1)^{2 s_i}}\,,
 \label{eq-d2NdtdE}
 \eeq
where $E^2= p^2 c^2 + m^2 c^4$, the greybody factor $\Gamma_{s_i}$ is a dimensionless absorption probability for the emitted species (it is in general a function of $E$, $M_{BH}(t)$ and the particle's internal dof and rest mass), and $g_i$ are the internal dof of the $i$-th particle, which account for spin, polarization and color. For the counting of the internal dof, we follow the notation of \cite{Arbey:2019mbc} (see their table 3).

Let us consider in some detail the SM. For the Higgs boson ($s=0$), $g_{h^0}=1$. For the massless ($s=1$) photon and the 8 gluons, $g_\gamma =2$ and  $g_g =16$. For the massive ($s=1$) W$^\pm$ and Z bosons, $g_{W^+} = g_{W^-} =g_{Z} = 3$. As for the fermions ($s=1/2$): the charged leptons, being Dirac fermions, have $g_e=g_\mu=g_\tau =4$; the neutrinos have $g_{\nu_e}=g_{\nu_\mu}=g_{\nu_\tau}=2(4)$ in the case they are Majorana (Dirac) particles, respectively; the quarks have $g_u=g_c=g_t=g_d=g_s=g_b=12$. Finally, one might also include the graviton ($s=2$), with $g_G=2$.

As $E \rightarrow +\infty$, each species approaches the geometrical optics limit
\beq
\Gamma_{s_i} (E, T_{BH}(t)) \rightarrow 
27 \frac{ E^2 (M_{BH}(t) c^2)^2}{(M_{Pl}  c^2)^4} = 
\frac{27}{(8 \pi)^2} \, \left( \frac{E}{k_B T_{BH}(t)} \right)^2 \,,
\label{eq-opt}
\eeq 
but falls off more quickly as $E \rightarrow 0$, with the higher spins producing the stronger cutoffs. When a massless particle scatters off a non-rotating, uncharged hole, the low-energy (that is $G M E/ \hbar c^3 \ll 1$) analytic form of $\Gamma_{s_i}$, averaged over all orientations of the hole with respect to the spin-weighted spherical harmonics and angular momentum quantum numbers of the incoming field, has been computed in ref.\,\cite{Starobinsky:1974} (see also \cite{Page:1976df}). The non-0 rest mass $m_{DM}$ of the DM particles acts as a cutoff in their emission at low energy, but the precise shape of the greybody factor around this cutoff is not relevant here since $k_B T_{BH} \gg m_{DM}c^2$ at all times.

\subsection{Rate of mass loss and BH lifetime}

The rate of mass loss for an evaporating BH is proportional to the total power emitted
\beq
-c^2 \frac{dM_{BH}}{dt} =  \frac{dE}{dt} 
= \sum_i  \int_0^\infty  dE \, E\,\frac{d^2N_i}{dt\, dE} \, .
\label{eq-Mloss}
\eeq
To parametrize this, Page \cite{Page:1976df} introduced the adimensional 
Page function $f(M_{BH})$ such that
\beq
\frac{dM_{BH}}{dt} = - \frac{c^2 M_{Pl}^4}{\hbar} \frac{f(M_{BH})}{M_{BH}^2} \,,
\eeq
where the time dependence of $M_{BH}$ is understood.

The BH lifetime $\tau$ is
\beq
\frac{\tau}{\hbar} = \frac{1}{c^2 M_{Pl}^4 }  \int_0^{M_{BH}} dM  \frac{M^2}{f(M)}\,.
\eeq
Defining $t_{ev}$ as the time of the BH evanescence, we practically have $\tau=t_{ev}$.

For the SM, the function $f(M_{BH})$ is constant over the range of BH masses we are interested in, namely $10^{-5}-10^9$ g, because all SM dof are already radiated by a $10^9$ g BH. Its value (not including gravitons) is $f(M_{BH})= 4.26 \times 10^{-3}$. This is the value that we are going to use in the following.

Assuming $f(M_{BH})$ constant over the BH lifetime, the latter is easily found to be
\beq
\frac{\tau}{\hbar} = \frac{1}{3  f(M_{BH})} \frac{(M_{BH} c^2)^3}{(M_{Pl} c^2)^4} \, ,
\label{eq-tevBH}
\eeq
which corresponds to the following time dependence of the BH mass, 
\beq
M_{BH}(t) = M_{BH} \left( 1- \frac{t}{\tau}\right)^{1/3} \, .
\label{eq-MBHt}
\eeq

\section{Instantaneous primary spectrum}

In this paper we consider the possibility that, on top of the SM, a DM candidate of mass $m_{DM}$ is produced in the evaporation process.
It is well known that there are two possible solutions, denoted as ``light'' and ``heavy'' DM, according to the fact
that the particles are produced during all the BH lifetime or just in its final stages (see \emph{e.g.} \cite{Masina:2020xhk} and references therein). 
If $m_{DM} c^2 < k_B T_{BH}$, the DM candidate belongs to the ``light'' category, otherwise to the ``heavy'' one.
In this paper we focus on light DM. 

Eq.\,(\ref{eq-d2NdtdE}) gives the instantaneous spectrum of the particles of type $i$ emitted by a single BH. The maximum of the energy distribution is at $E \sim k_B T_{BH}$. For sufficiently light DM,  the ultra relativistic limit, $E \approx p c > m_{DM} c^2$, is thus justified. As the BH evaporates, its temperature increases, and the relativistic limit is satisfied \emph{a fortiori}.
In the relativistic limit, the instantaneous distribution of emitted momentum, per dof, is
\beq
\frac{d^2 N}{ dt\,d(c p(t))} (c p(t),T_{BH}(t)) 
=\frac{1}{2 \pi \hbar } \, \Gamma_{s}(cp(t),T_{BH}(t)) \,  \frac{1}{e^\frac{cp(t)}{k_B T_{BH}(t)} - (-1)^{2 s}}\,,
 \label{eq-d2Ndtdcp}
\eeq
where we introduced explicitly the time dependence for the sake of clarity and $s$ is the spin of DM.

To obtain the number densities (per dof) of the emitted particles, one has to multiply the above expression 
by the number density of the BHs $n_{BH}(t)$
\beq
\frac{d^2 n}{ dt\,d(c p(t))} (c p(t),T_{BH}(t))= n_{BH}(t)\, \frac{d^2 N}{ dt\,d(c p(t))} (c p(t),T_{BH}(t))\,.
\label{eq-d2ndtdcp}
\eeq

\subsection{Instantaneous distribution: geometrical optics approximation}

Using eq.\,(\ref{eq-d2Ndtdcp}) with the geometrical optics limit, eq.\,(\ref{eq-opt}), 
at a fixed time $t$, the instantaneous distribution (per particle dof) has the form
\beq
\frac{d^2 N}{ dt\,d(c p(t))} (c p(t),T_{BH}(t))
= \frac{1}{2 \pi \hbar } \,\frac{27}{(8 \pi)^2}\, \left(  \frac{c p(t) }{ k_B T_{BH}(t)} \right)^2  \frac{ 1 }{e^\frac{c p(t)}{k_B T_{BH}(t)} - (-1)^{2s}} \, .
\label{eq-d2Ndtdcp-GO}
 \eeq

\begin{figure}[h!]
\vskip .0 cm 
 \begin{center}
 \includegraphics[width=7.6cm]{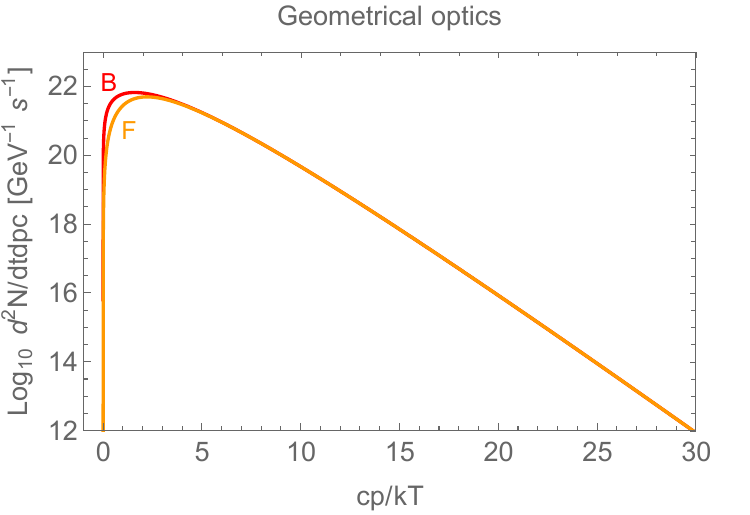}   \,\,\,
 \includegraphics[width=7.6cm]{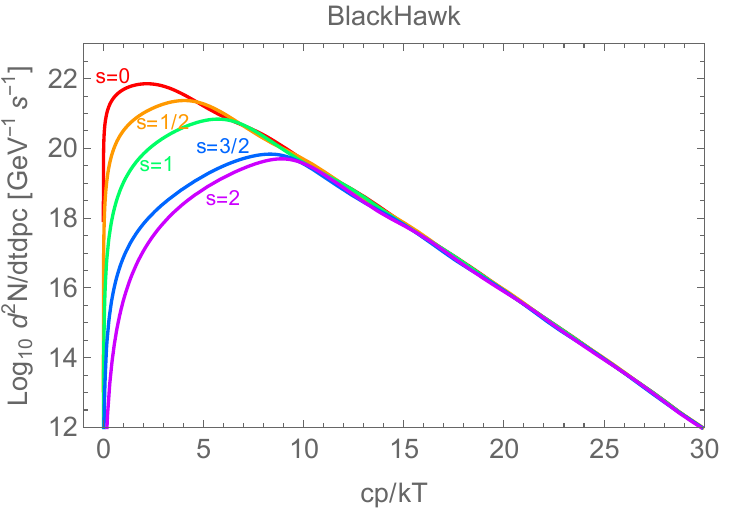}   
 \end{center}
\caption{\baselineskip=15 pt \small  
Left: instantaneous distribution as a function of $x(t_f)$ in the geometrical optics approximation. 
Right: the same using the numerical results from \texttt{BlackHawk}.}
\label{fig-inst-distr}
\vskip .2 cm
\end{figure}

Consider for definiteness the formation time $t_f$.  
It is useful to introduce the quantity
\beq
x(t_f) \equiv \frac{c p(t_f)}{k_B T_{BH}} \, ,
\label{eq-xtf}
\eeq
as it allows to get rid of the BH mass dependence. Indeed, the instantaneous distribution becomes
\beq
\left. \frac{d^2 N}{ dt\,d(c p(t))} (c p(t),T_{BH}(t))  \right|_{t_f}
= \frac{1}{2 \pi \hbar }  \, \frac{27}{(8 \pi)^2}  \,  x(t_f)^2 \,  \frac{ 1 }{e^{x(t_f)} - (-1)^{2s}}\,.
\eeq
In the left plot of fig.\,\ref{fig-inst-distr}, we show the instantaneous momentum distribution, 
as a function of $x(t_f)$, for bosons (B) and fermions (F) respectively, in the geometrical optics approximation.

\subsection{Instantaneous distribution: numerical results from \texttt{BlackHawk}}

This has to be confronted with the corresponding numerical results obtained using \texttt{BlackHawk} \cite{Arbey:2019mbc}, which provides the instantaneous momentum distribution
at various times, from formation to evaporation. \texttt{BlackHawk} has been modified to obtain the additional DM spectra, a modification already highlighted in the manual and which will be incorporated in a future version of the public code. In particular, greybody factors for the spin $3/2$ case have been computed. For a direct comparison with the geometrical optics approximation, the distributions in the right panel of fig.\,\ref{fig-inst-distr} are normalized per particle dof.
We can see that the $s=0$ case is quite similar with respect to the geometrical optics limit for the boson, while the $s=1/2$ case is a bit suppressed with respect to the geometrical optics limit for the fermion. The higher the spin is, the more the distribution is suppressed at low energies, so that the mean momentum gets higher.
 
\begin{figure}[h!]
\vskip .0 cm 
 \begin{center}
\includegraphics[width=7.6cm]{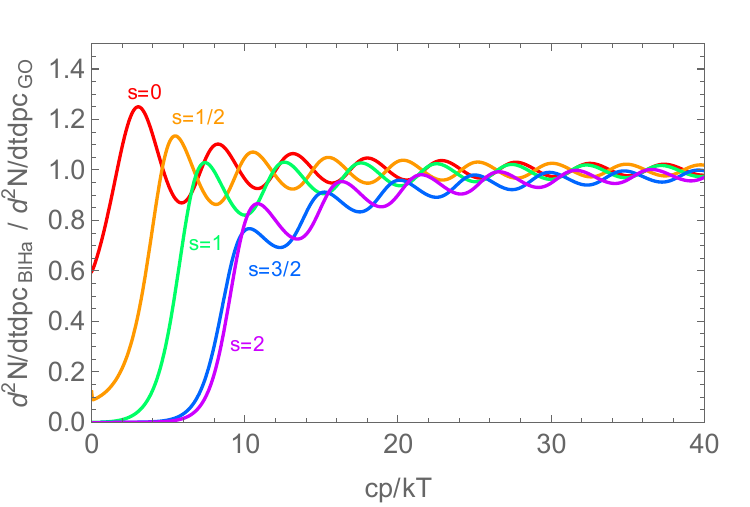}   
 \end{center}
\caption{\baselineskip=15 pt \small  
Ratio of the instantaneous distributions at formation as given by \texttt{BlackHawk} over the corresponding geometrical optics limit.}
\label{fig-inst-distr-f-ratio}
\vskip .2 cm
\end{figure}

A more precise comparison with the geometrical optics approximation can be established by studying the ratio of the instantaneous distributions at formation for \texttt{BlackHawk} over the corresponding geometrical optics limit, as shown in fig.\,\ref{fig-inst-distr-f-ratio}. As is well known (see \emph{e.g.} \cite{Page:1976df} and references therein), the ratio for $s=0$ and $s=1/2$ at low energy is a constant, respectively equal to $16/27$ and $2/27$; the figure correctly reproduces this low energy behavior. 

\section{From formation to evaporation}

Let us define $f(t)\equiv\rho_{BH}(t)/\rho_R(t) \propto a(t)$. 
As time increases, it is then possible that BHs come to dominate the energy content of the universe before they completely evaporate \cite{Barrow:1990he, Baumann:2007yr, Fujita:2014hha}:
this situation is referred to as BH domination. The scenario of evaporation before this happens is referred to as radiation domination.

\subsection{Radiation vs BH domination}

\begin{figure}[h!]
\vskip .0 cm 
 \begin{center}
 \includegraphics[width=7.6cm]{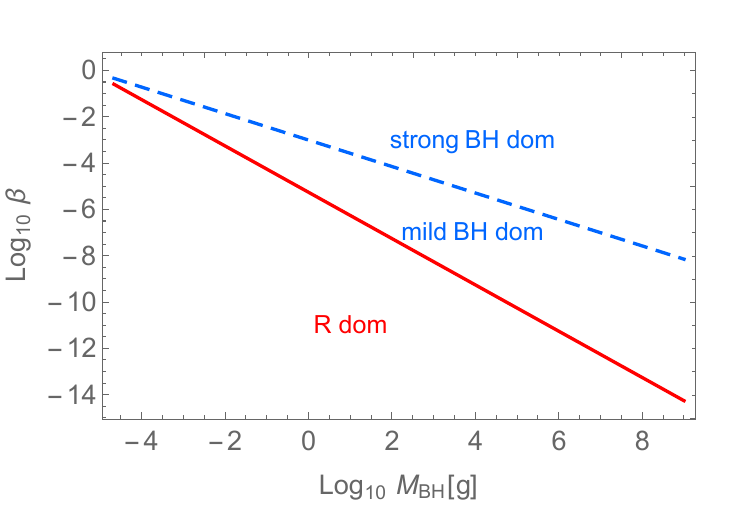}  \,\,  \includegraphics[width=7.6cm]{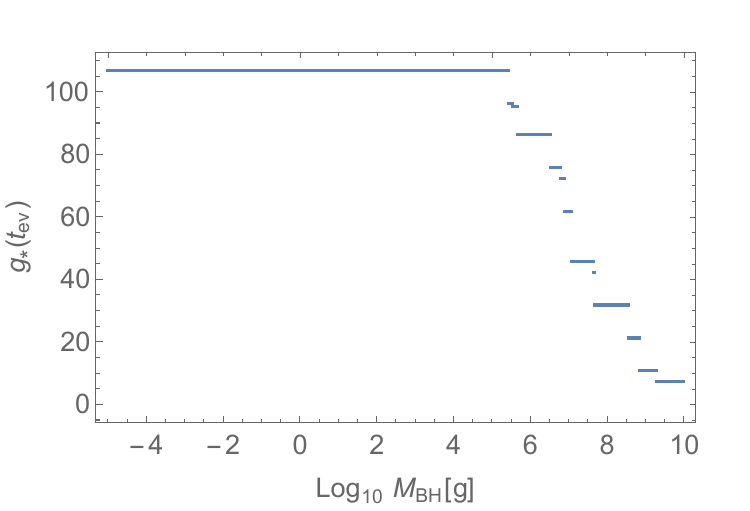} 
 \end{center}
\caption{\baselineskip=15 pt \small  
Left: the solid (red) line is $\bar \beta$, and the dashed (blue) line separates regions where the change in the scale factor during BH domination period is smaller (mild) or greater (strong) than the change during the radiation domination period. Right: $g_{*}(t_{ev})$ for radiation domination (clearly, this is an approximation by a step function, while the real function $g_*(t_{ev})$ is a smooth continuous function).}
\label{fig-bbeta}
\vskip .2 cm
\end{figure}

\noindent {\bf Radiation domination} \\
The BHs evaporate during the radiation epoch if ${f(t_{ev})}\lesssim 1$. We define $\bar \beta$ the maximum value of $\beta$ corresponding to radiation domination, namely the value of $\beta$ leading to ${f(t_{ev})} \simeq 1$;
this value can be obtained from the following relation
\beq
 {\bar \beta} \simeq \frac{f(t_f)}{f(t_{ev})} = \frac{a(t_f)}{a(t_{ev})} = \left( \frac{t_f}{t_{ev}} \right)^{1/2}
= \left(     \frac{ 3 f(M_{BH})}{\gamma}  \right)^{1/2}  \frac{M_{Pl} } {M_{BH} }   \, ,
\eeq
where we used eqs.\,(\ref{eq-tfBH}) and (\ref{eq-tevBH}) to obtain the last equality. The value of $\bar \beta$ as a function of the BH mass is shown in fig.\,\ref{fig-bbeta}, 
taking for definiteness $f(M_{BH})$ as in the SM and $\gamma=0.2$. For all the values of $\beta \lesssim \bar \beta$ (red solid line), the primordial BHs evaporate before they come to dominate the energy content of the universe and, also in this case, the increase in the scale factor is
\beq
\frac{a(t_f)}{a(t_{ev})} = {\bar \beta} \label{eq-betaR} \, .
\eeq

\noindent {\bf BH domination}\\
In the case of BH domination, we have instead to consider: first, the radiation dominated period from the formation time, $t_f$, to the time when BHs start to dominate, $t_{BH}$, such that $f(t_{BH})\simeq 1$; and second, the matter dominated period from $t_{BH}$ to $t_{ev}$. The first period is characterized by the following increase in the scale factor
\beq
{\beta} \simeq \frac{f(t_f)}{f(t_{BH})} = \frac{a(t_f)}{a(t_{BH})} = \left( \frac{t_f}{t_{BH}} \right)^{1/2} \, .
\eeq
The second period is characterized by
\beq
\frac{a(t_{BH})}{a(t_{ev})} = \left( \frac{t_{BH}}{t_{ev}} \right)^{2/3} =  \left( \frac{1}{\beta^2} \frac{t_{f}}{t_{ev}} \right)^{2/3}  
 =\frac{1}{\beta^{4/3}}   \left(     \frac{ 3 f(M_{BH})}{\gamma}  \right)^{2/3}       \left(   \frac{M_{Pl} } {M_{BH} }      \right)^{4/3} 
 =\left( \frac{\bar \beta}{\beta}  \right)^{4/3}\, .
\eeq
Putting together
\beq
\frac{a(t_f)}{a(t_{ev})} = \frac{a(t_f)}{a(t_{BH})} \frac{a(t_{BH})}{a(t_{ev})} 
=   \frac{1}{\beta^{1/3}}   \left(     \frac{ 3 f(M_{BH})}{\gamma}  \right)^{2/3}       \left(   \frac{M_{Pl} } {M_{BH} }      \right)^{4/3} 
= \frac{\bar \beta^{4/3}}{\beta^{1/3}}  \,.
\eeq

Notice also that the two periods have the same increase in the scale factor, $\frac{a(t_f)}{a(t_{BH})} = \frac{a(t_{BH})}{a(t_{ev})}$, 
if $ \beta = \bar \beta^{4/7}$, so that $ t_{BH}= t_f /\bar \beta^{8/7}$.
We can define as mild and strong BH domination the regions where the increase in the scale factor in the first period is respectively larger and smaller than the second one, as shown in fig.\,\ref{fig-bbeta}. Notice that in the very strong BH domination region (close to $\beta \sim 1$) one can neglect the first period of radiation domination, having
\beq
\frac{a(t_f)}{a(t_{ev})} =   \left(     \frac{ 3 f(M_{BH})}{\gamma}  \right)^{2/3}       \left(   \frac{M_{Pl} } {M_{BH} }      \right)^{4/3}  
\sim \bar \beta^{4/3}\, . \label{eq-betaBH}
\eeq

\subsection{Radiation temperature at evaporation}

For radiation domination, combining eqs.\,(\ref{eq-F1}) and (\ref{eq-raddom}), we have
\beq
\frac{8 \pi G}{3} \rho_R(t_{ev})  = \frac{1}{4 \tau^2} \,.
\eeq
Using also eqs.\,(\ref{eq-rhoRT}) and (\ref{eq-tevBH}), we obtain
\beq
k_B T_R(t_{ev}) = 
 \left(   3 f(M_{BH}) \right)^{1/2}  
 \left( \frac{ 45  }{16 \pi^3 g_{*}(t_{ev})}     \right)^{1/4}  \left( \frac{     M_{Pl} }{  M_{BH}  }  \right)^{3/2} (M_{Pl} c^2)\,.
\label{eq-TevR} 
\eeq
The values of $g_{*}(t_{ev})$ as a function of the BH mass are shown in the right panel of fig.\,\ref{fig-bbeta}.

For full BH domination, we can grossly assume that all the energy stored in the BH density goes, after their evaporation, into the radiation energy density of the (SM and possibly beyond SM) particles emitted by the BHs. These particles rapidly thermalize as soon as they are emitted, so that
$\rho_R(t_{ev}^+) \approx \rho_{BH}(t_{ev}^-)$.
Combining eqs.\,(\ref{eq-F1}) and (\ref{eq-matdom}), we have
\beq
\frac{8 \pi G}{3} \rho_{R}(t_{ev}^+)=\frac{8 \pi G}{3} \rho_{BH}(t_{ev}^-)  = \frac{4}{9 \tau^2} \, .
\eeq
We then have
\beq
\frac{T_R^R(t_{ev})}{T_R^{BH}(t_{ev}^+)} = \left( \frac{9}{16} \right)^{1/4}  \, .
\label{eq-TevBH}
\eeq
For BH domination, the radiation temperature after evaporation gets slightly enhanced with respect to radiation domination, the difference is about $15\%$.

\section{Distribution at evaporation}

The distribution of DM momentum at evaporation $F(c p(t_{ev}), t_{ev})$ is a superposition of all the instantaneous distributions, each redshifted appropriately from its time of emission $t_{em}$ (see \emph{e.g.} ref.\,\cite{Bugaev:2000bz})
\beq
F(c p(t_{ev}),t_{ev}) 
=\int_{t_{em}}^{t_{ev}} dt \,  \frac{d^2 N}{ dt\,d(c p(t)) }  \left( \underbrace{c p(t_{ev}) \frac{a(t_{ev})}{a(t)} }_{cp(t)}, T_{BH}(t) \right) \,\,  \frac{a(t_{ev})} {a(t)} \, .
\label{eq-Fev}
\eeq
Notice that $t_{em}$ might be larger than $t_f$ if the initial BH temperature is smaller than the particle mass. Since we are interested in light DM, we have $t_{em}= t_f$.

For radiation domination from formation to evaporation, the ratio of scale factors in eq.\,(\ref{eq-Fev}) is
\beq
 \frac{a(t_{ev})}{a(t)} = \left( \frac{t_{ev}}{t} \right)^{1/2} \,.
 \eeq

For BH domination, the integral of eq.\,(\ref{eq-Fev}) should be split into two contributions, corresponding to a first period of radiation domination, and a second of BH domination
\beq
F(c p(t_{ev}),t_{ev}) = F^R(c p(t_{ev}),t_{ev}) + F^{BH}(c p(t_{ev}),t_{ev})  \,.
\eeq
For the second period of BH domination, starting at $t_{BH}=t_f/\beta^2$ and ending at $t_{ev}$, 
the ratio of scale factors to be put in the integrand is
\beq
 \frac{a(t_{ev})}{a(t)} = \left( \frac{t_{ev}}{t} \right)^{2/3} \,. 
\eeq
For the first period of radiation domination, starting at $t_f$ and ending at $t_{BH}$,
the ratio of scale factors to be put in the integrand is rather
\beq
 \frac{a(t_{ev})} {a(t)} =  \frac{a(t_{ev})} {a(t_{BH})} \frac{a(t_{BH})} {a(t)} 
 = \left( \frac{t_{ev}}{t_{BH}} \right)^{2/3} \left( \frac{t_{BH}}{t} \right)^{1/2}  \, .
\eeq
For very strong BH domination the dominant contribution comes from $F^{BH}$.

In order to get rid of the BH mass dependence, it is useful to define (as suggested by Lennon \emph{et al.} in ref.\,\cite{Lennon:2017tqq}) the adimensional momentum (notice that it is not the same as in eq.\,(\ref{eq-xtf}))
\beq
x(t_{ev}) \equiv \frac{c p(t_{ev})}{k_B T_{BH} } \, ,
\eeq
with the related adimensional momentum distribution at evaporation
\beq
\tilde F(x(t_{ev}))  \equiv \frac{(k_B T_{BH})^3}{(M_{Pl} c^2)^2} F(c p(t_{ev}),t_{ev})  \,.
\label{eq-tFev}
\eeq

\subsection{Distribution at evaporation: geometrical optics approximation}
 
In the geometrical optics limit of eq.\,(\ref{eq-d2Ndtdcp-GO}), eq.\,(\ref{eq-Fev}) becomes
\beq
F(c p(t_{ev}),t_{ev}) 
= \frac{27}{2 \pi \hbar }    \frac{ 1}{(M_{Pl}  c^2)^4} 
\int_{t_{f}}^{t_{ev}} dt \,     { (M_{BH}(t) c^2)^2} \frac{ (c p(t_{ev}) 
\frac{a(t_{ev})}{a(t)})^2 }{e^\frac{ c p(t_{ev}) \frac{a(t_{ev})}{a(t)}}{k_B T_{BH}(t)} - (-1)^{2s}}    \,\,  \frac{a(t_{ev})} {a(t)} \,.
\eeq

We derive simplified analytical expressions in the case that $f(M_{BH})$ is constant over the BH lifetime: using eqs.\,(\ref{eq-TBH}), (\ref{eq-tevBH}) and (\ref{eq-MBHt}), we obtain
\beq
F(cp(t_{ev}),t_{ev}) =  ( c p(t_{ev})   )^2  \frac{9}{2 \pi  }  { \frac{1}{ f(M_{BH})} \frac{(M_{BH} c^2)^5}{(M_{Pl} c^2)^8} }\, I(x(t_{ev})) \,,
\eeq
where for full radiation domination
\beq
I_R(x(t_{ev}))= \int_{t_{f}/t_{ev}}^{1} dy  \,  \frac{ (1-y)^{2/3} }{y^{3/2}}  \frac{ 1} {  e^ { x(t_{ev}) \frac{ (1-y)^{1/3} }{ y^{1/2}} } - (-1)^{2s}    }  \,  ,
\eeq
while for full BH domination
\beq
{I_{BH}(x(t_{ev}))} =
\int_{t_{f}/t_{ev}}^{1} dy  \,  \frac{ (1-y)^{2/3} }{y^{2}}  \frac{ 1} {  e^ { x(t_{ev}) \frac{ (1-y)^{1/3} }{ y^{2/3}} } - (-1)^{2s}    }   \,  .
\eeq

\begin{figure}[h!]
\vskip .0 cm 
 \begin{center}
\includegraphics[width= 7.8 cm]{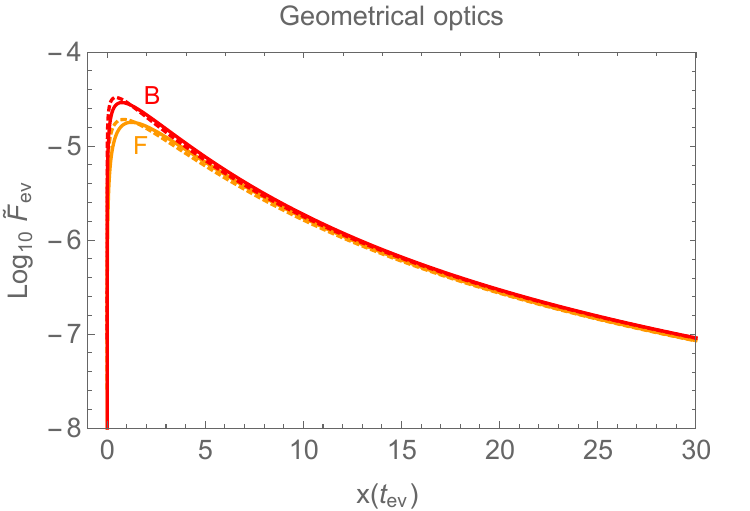}   
\,\,  \includegraphics[width= 7.8 cm]{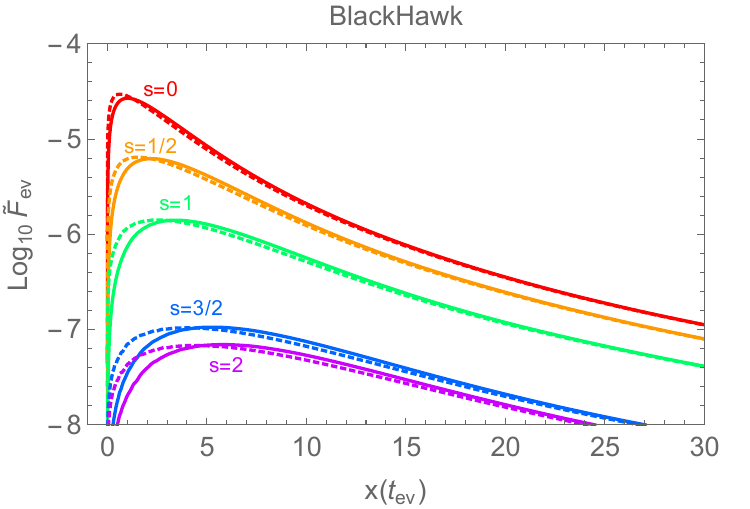}   
\end{center}
\caption{\baselineskip=15 pt \small  
Left: Solid (dotted) lines: Momentum distribution at evaporation including redshift and assuming radiation (BH) domination,
in the geometrical optics approximation. 
Right: Solid (dotted) lines: Momentum distribution at evaporation including redshift and assuming radiation (BH) domination, calculated with \texttt{BlackHawk}, for various spins, and normalized to the number of dof.}
\label{fig-distr-ev}
\vskip .2 cm
\end{figure}

The adimensional momentum distribution at evaporation, eq.\,(\ref{eq-tFev}), becomes
\beq
\tilde F(x(t_{ev}))  
= \frac{1}{(8 \pi)^5}  \frac{9}{2 \pi  }  { \frac{1}{ f(M_{BH})}  }  \,x(t_{ev})^2 \, I(x(t_{ev})) \,.
\label{eq-tFev-GO}
\eeq
We show this quantity in the left panel of fig.\,\ref{fig-distr-ev} for the two scenarios of radiation domination with $\beta = \bar \beta$ (solid) and for full BH domination (dotted).  Notice that the difference between the two scenarios is quite small. To reproduce the case of radiation domination with a different value of $\beta$, one has just to suppress $\tilde F$ by the factor $\beta/\bar \beta$.
These results agree with the ones of Lennon \emph{et al.} \cite{Lennon:2017tqq}.

\subsection{Distribution at evaporation: numerical results from \texttt{BlackHawk}}

The quantity $\tilde F$ derived from \texttt{BlackHawk} is shown in the right panel of fig.\,\ref{fig-distr-ev}, 
assuming radiation domination with $\beta = \bar \beta$ (solid) and full BH domination (dotted). The suppression due to the different values of the spin is manifest. The case $s=0$ is quite similar to the bosonic case of the geometrical optics limit. For higher spins, the geometrical optics approximation becomes worser. Again, we can see that the difference between radiation domination with $\beta=\bar \beta$ and full BH domination is quite small.

\section{The DM phase space distribution}

We now calculate the DM phase space distribution, as it is the essential ingredient to derive both the DM abundance and, using the publicly available code \texttt{CLASS}, 
the transfer function for structure formation. The DM phase space distribution (psd) per dof, $f_{DM}$, at time $t$, is defined as
\beq 
  f_{DM}(p,t)  \equiv \frac{1}{g_{DM}} \frac{d n_{DM}}{d^3 (cp)} 
=\frac{1}{g_{DM}}   \frac{1}{(cp)^2 d\Omega} \frac{d n_{DM}}{ d(cp) }\,,
\label{eq-defpsd}
\eeq
where $g_{DM}$ is the number of DM dof, $n_{DM}$ the DM number density (scaling as $a^{-3}$ from evaporation time), $p$ is the DM momentum (scaling as $a^{-1}$) and $d\Omega = 4\pi$ is the solid angle.

For DM produced by evaporating BHs, by using eqs.\,(\ref{eq-d2ndtdcp}) and (\ref{eq-Fev}), we obtain that the psd at time $t_{ev}$ is
\beq  
f_{DM}(p(t_{ev}),t_{ev}) \,d\Omega =  \frac{1}{(cp(t_{ev}))^2 }\, n_{BH}(t_{ev})   \,F(cp(t_{ev}),t_{ev})
 =\,n_{BH} (t_{ev})\, \frac{(M_{Pl} c^2)^2}{(k_B T_{BH})^5} \,\, {\frac{ \tilde F(x(t_{ev})) }{x(t_{ev})^2}}\,,
\eeq
where in the last equality we switched to the adimensional quantity $\tilde F$ defined in eq.\,(\ref{eq-tFev}).
The BH number density at the time of evaporation is related to the one at formation by
\beq
n_{BH} (t_{ev}) = n_{BH} (t_f) \left( \frac{a(t_f)}{a(t_{ev})}\right)^3 \,.
\label{eq-nBHtev}
\eeq
Recalling the definition of $\beta$ in eq.\,(\ref{eq-defbeta}) we have
\beq
n_{BH} (t_f) = \beta \,\frac{\rho_{R}(t_f)}{M_{BH}} 
=\beta \, 3 (4 \pi \gamma)^2 \left( \frac{k_B T_{BH}}{\hbar c} \right)^3 \,,
\label{eq-nBHtf}
\eeq
where the last equality has been obtained by combining eq.\,(\ref{eq-MPBH}) for the BH mass at the formation time
and the Friedman equation (\ref{eq-F1}).
By exploiting the last two equations, we have
\beq
(\hbar c)^3\,f_{DM}(p(t_{ev}),t_{ev}) \, d\Omega  = \beta \, 3 (4 \pi \gamma)^2    \,\left( \frac{a(t_f)}{a(t_{ev})}\right)^3 \, \frac{(M_{Pl} c^2)^2}{(k_B T_{BH})^2} \, \frac{\tilde F(x(t_{ev}))}{x(t_{ev})^2}  \,.
\eeq
Using also eqs.\,(\ref{eq-betaR}) and \eqref{eq-betaBH}, we finally have\footnote{
In order to match our expression with the results of Baldes \emph{et al.} \cite{Baldes:2020nuv}, we define
$\xi= \frac{(M_{Pl} c^2)^2}{(k T_{BH})^2} = (8 \pi)^2  \left( \frac{ M_{BH} }{ M_{Pl} } \right)^2$.
Using eqs.\,(\ref{eq-betaR}) and \eqref{eq-betaBH}, we obtain that, at the evaporation time
$$
(\hbar c)^3  f_{DM}(p(t_{ev}),t_{ev}) \, d\Omega 
= 
\zeta_{R/BH} \,  \, 
\frac{ \xi \tilde F(x(t_{ev}))}{x(t_{ev})^2}\,,
$$
where
$$
\zeta_{R} = \beta \, 3 (4 \pi)^2 \gamma^{1/2}     \,   (3 f(M_{BH}))^{3/2}     \left(    \frac{M_{Pl} } {M_{BH} }   \right)^3
\, , \quad
\zeta_{BH} = \, 3 (4 \pi)^2     (3 f(M_{BH}))^{2}            \left(    \frac{M_{Pl} } {M_{BH} }   \right)^4 \, ,
$$
which is the same as eq.\,(3.8) of Baldes \emph{et al.} \cite{Baldes:2020nuv}, apart from the fact that they have a factor of 4 instead of 3 for BH domination.
}
\beq
(\hbar c)^3  f_{DM}(p(t_{ev}),t_{ev}) \, =
{A_{R,BH}}     \,   \, \frac{\tilde F(x(t_{ev}))}{x(t_{ev})^2}\, ,
\label{eq-psdDM}    
\eeq
with
\beq
A_R=  \beta \, 3 (8 \pi)^2  (4 \pi) \gamma^{1/2}    \,    (3 f(M_{BH}))^{3/2}  \left(    \frac{M_{Pl} } {M_{BH} }   \right) \, , \quad
A_{BH} = \, 3 (8 \pi)^2  (4 \pi)    \,  (3 f(M_{BH}))^{2}   \left(    \frac{M_{Pl} } {M_{BH} }   \right)^2  \,.
 \label{eq-psdDMA}   
\eeq

In order to proceed further in the calculation of the DM abundance and transfer function, it is useful to establish a comparison with the well known case of a thermal DM candidate.

 \subsection{Psd for thermal DM}

An hypothetical thermal DM decoupling a $t=t_{d}$ would have the following psd (per dof)
\beq
(\hbar c)^3  f^{th}_{DM} (p(t_{d}),t_{d}) =   
\frac{1}{ (2 \pi)^3}  \, \frac{ 1 }{e^\frac{c p(t_{d})}{k_B T_{DM}(t_{d})} - (-1)^{2s}} \,,
\eeq
where $T_{DM}(t_{d})$ is the temperature of the DM at decoupling, which for a thermal DM candidate has to be identified with the temperature of the radiation bath from which it decouples, namely $T_{DM}(t_{d})= T_R(t_{d})$.

At later times, $t>t_d$, both $p(t)$ and $T_{DM}(t)$ scale as the inverse of the scale factor.
It is then useful to define the time independent parameter 
\beq
q \equiv q(t) =  \frac{c p(t)}{k_B T_{DM}(t)} \,,
\label{eq-defq}
\eeq  
and re-express the psd in terms of $q$
\beq
(\hbar c)^3  f^{th}_{DM} (q) =   
\frac{1}{ (2 \pi)^3}  \, \frac{ 1 }{e^q - (-1)^{2s}} \,.
\label{eq-psdth}
\eeq

It is also useful to express the DM temperature now, $T_{DM}(t_0)$, in units of the photon temperature now, $T_\gamma (t_0)\simeq 2.7$ K.

If decoupling happened just before recombination, $t_d=t^-_r$ (or more generally when, in addition to DM, only photons, neutrinos and electrons were relativistic), clearly $T_{DM}(t^-_r)=T_{\nu}(t^-_r)=T_{\gamma}(t^-_r)$. Because of the successive reheating of the photons at recombination, at the present time $t_0$ we have $T_{DM}(t_0)=T_{\nu}(t_0) = \left(\frac{4}{11}\right)^{1/3} T_\gamma (t_0)$.

If decoupling happened much earlier, when also other SM particles were relativistic, allowing for a possible entropy non conservation from decoupling to recombination, $\bar \alpha (s a^3)_d= (s a^3)_{r}$, we have
\beq
\frac{a(t_r)}{a(t_d)} = {\bar \alpha}^{1/3}  \left(  \frac{g_{*,S} (t_d)}{ g_{*,S} (t_r)} \right)^{1/3}   \frac{T_R(t_d)}{T_R(t_r)}
=  {\bar  \alpha}^{1/3} \left(  \frac{g_{*,S} (t_d)}{ g_{*,S} (t_r)} \right)^{1/3}   \frac{T_R(t_d)}{T_\nu(t_r)}\,,
\eeq
where $g_{*,S}(t_d) = 106.75$ for the complete SM, $g_{*,S}(t_r) = 
10.75$ (including photons, neutrinos and electrons). The DM temperature instead simply scales as the inverse of the scale factor
\beq
\frac{T_{DM}(t_d)}{T_{DM}(t_r)} =\frac{T_{R}(t_d)}{T_{DM}(t_r)} = \frac{a(t_r)}{a(t_d)} \,.
\eeq
Hence, the DM particles do not share the entropy release from the successive annihilations and their temperature is suppressed at recombination by a factor
\beq
  \frac{T_{DM}(t_r)}{T_\nu(t_r)}=   \frac{1}{\bar \alpha^{1/3}}   \left(  \frac{g_{*,S} (t_r)}{ g_{*,S} (t_d)} \right)^{1/3} <1  \, .
\eeq
Since from recombination to now both $T_{DM}$ and $T_\nu$ scale as the inverse of the scale factor, the same relation
holds today
\beq
\frac{T_{DM}(t_0)}{T_\nu(t_0)} =  \frac{1}{\bar \alpha^{1/3}}   \left(  \frac{g_{*,S} (t_r)}{ g_{*,S} (t_d)} \right)^{1/3} 
<1  \, .
\eeq

Notice that \texttt{CLASS} actually uses the DM temperature now (or rather the temperature of the radiation bath at decoupling, rescaled to its value now) in units of the photon temperature now. For an early decoupled particle then
\beq
\frac{ T_{DM}(t_0)}{T_\gamma(t_0)} =\frac{T_{DM}(t_0)}{T_\nu(t_0)} \frac{T_{\nu}(t_0)}{T_\gamma(t_0)}  
= \frac{1}{\bar \alpha^{1/3}}   \underbrace{ \left(  \frac{g_{*,S} (t_r)}{ g_{*,S} (t_d)} \right)^{1/3}  }_{0.465} 
\underbrace{ \left(\frac{4}{11}\right)^{1/3}}_{0.714} \,\, \approx 0.332 \, ,
\label{eq-edec}
\eeq
where for the last approximation we used $\bar \alpha=1$, $g_{*,S} (t_d)=106.75$ and $g_{*,S} (t_r)=10.75$.

\subsection{Psd for BHs as a function of q}

DM particles from BHs were never in thermal equilibrium. Nevertheless we can imagine to deal with their distribution as they were ``decoupling'' at $t_{ev}$,  so that they would have the decoupling temperature $T_{DM}(t_{ev})= T_R(t^+_{ev})$,
which has already been derived, see eqs.\,(\ref{eq-TevR}) and (\ref{eq-TevBH}).

The time independent quantity $q$, defined in eq.\,(\ref{eq-defq}), is thus
\beq
q= \frac{c p(t_{ev})}{k_B T_{DM}(t_{ev})} =\frac{c p(t_{ev})}{k_B T_{R}(t^+_{ev})} 
= {\frac{T_{BH}}{T_R(t^+_{ev})}} \frac{c p(t_{ev})}{k_B T_{BH}}  \equiv {\alpha} \, x(t_{ev}) \,,
\label{eq-defqBH}
\eeq 
where, for radiation domination
\beq
{\alpha}_R 
= \frac{1}{8 \pi} 
 \frac{1}{ (   3 f(M_{BH}) )^{1/2}  }
 \left( \frac{16 \pi^3 g_{*}(t_{ev})}   { 45  }  \right)^{1/4}  \left( \frac{     M_{BH} }{  M_{Pl}  }  \right)^{1/2} \, ,
\label{eq-alfa}
\eeq  
while, according to eq.\,(\ref{eq-TevBH}), for BH domination $\alpha_{BH}= (9/16)^{1/4} \alpha_R$.

In terms of $q$, the psd of eq.\,(\ref{eq-psdDM}) then becomes\footnote{For comparison with previous literature,
notice that Baldes \emph{et al.} \cite{Baldes:2020nuv} assume $T_{DM}(t_{ev})=T_{BH}$, so that 
$$
q^B \equiv x(t_{ev})  =\frac{c p(t_{ev})}{k_B T_{BH} } 
\, .
$$
}
\beq
(\hbar c)^3  f_{DM}(q) \, 
=
 \, A_{R,BH}   \, \frac{\tilde F(q/\alpha)}{ (q/\alpha)^2}\,.
    \label{eq-psdDMq} 
\eeq
with $A_{R,BH}$ given by eq.\,(\ref{eq-psdDMA}). 
Since the evaporation process happens at early times, eq.\,(\ref{eq-edec}) applies also in this case, with the substitution $g_{*,S}(t_d) \rightarrow g_{*,S}(t_{ev})$.

\begin{figure}[h!]
\vskip .0 cm 
\begin{center}
\includegraphics[width=7.8 cm]{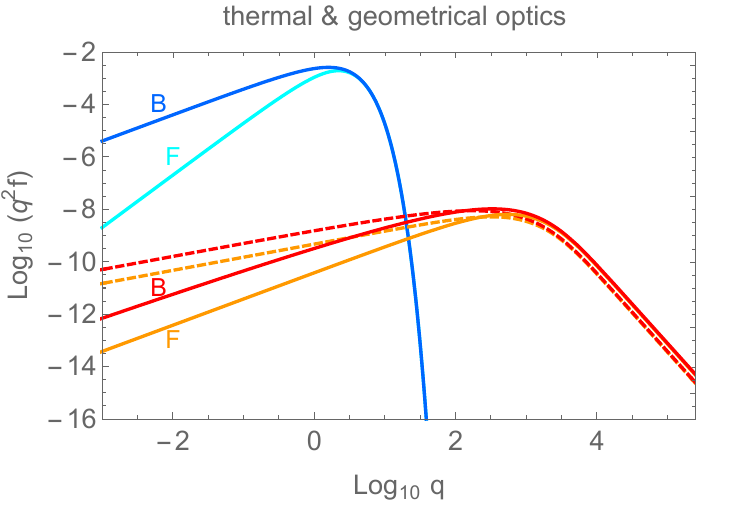}   \,\,  \includegraphics[width=7.8 cm]{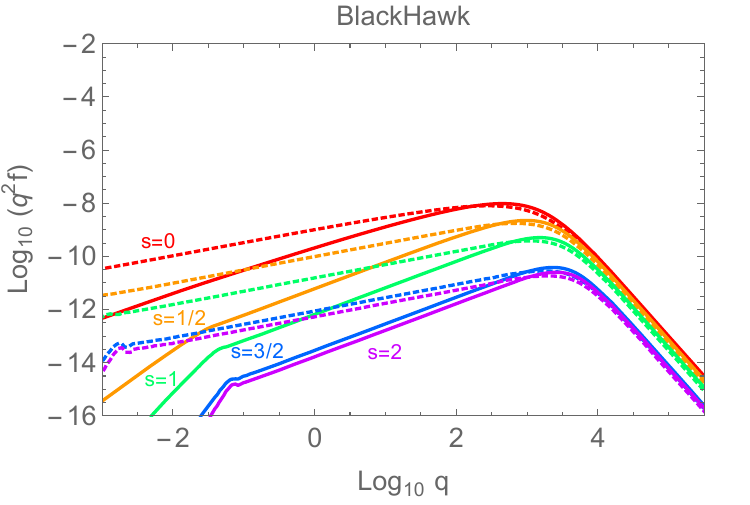} 
\end{center}
\caption{\baselineskip=15 pt \small  
Phase space distributions for $M_{BH}=1$ g. Left: R domination with $\beta=\bar \beta$ (solid) and full BH domination (dashed) with $M_{BH}=1$ g, in the geometrical optics approximation. Also shown is the thermal psd with $g_i=1$.
Right: psd for R domination with $\beta=\bar \beta$ (solid) and full BH domination (dashed) with $M_{BH}=1$ g, calculated with \texttt{BlackHawk}. }
\label{fig-psd}
\vskip .2 cm
\end{figure}

In the right panel of fig.\,\ref{fig-psd} we show the psd obtained from \texttt{BlackHawk}, taking $M_{BH}=1$ g, for radiation domination with $\bar \beta$ (solid) and full BH domination (dashed). For radiation domination with other values of $\beta$, the psd has to be suppressed by a factor $\beta/\bar \beta$. This has to be directly compared with the same situation, calculated in the geometrical optics approximation, 
as shown in the left panel. The suppression in the psd associated to higher spins is manifest. In the left panel we also show the thermal distribution (\ref{eq-psdth}) with one dof, $g_i=1$. It is clear that the spectrum of particles emitted by a $1$ g BH is much harder than the thermal one.

The psd for other values of the BH masses can be easily reconstructed in the following way. If for $M_{BH}=1$ g the peak is at about $ Log_{10}\, q \sim 3$, in general it is at $Log_{10}\, q \sim 3  +  Log_{10}  ( \frac{ M_{BH} }{1 \,{\rm g}})^{1/2}$. In addition, the psd gets suppressed by the factor $( \frac{1 \,{\rm g}} {M_{BH} })^{1/2}$.

\section{Dark Matter abundance}

The present  abundance of a stable DM particle produced by evaporation is directly related to the number-to-entropy density of such particles, $Y_{DM}(t_0) = {n_{DM}(t_0)}/{s(t_0)}$,
\beq
\Omega_{DM} 
= \frac{\rho_{DM}(t_0)}{\rho_c} = \frac{m_{DM}}{\rho_c} \frac{n_{DM}(t_0)}{s(t_0)}  \, s(t_0) \, ,
\label{eq-ODMg}
\eeq
where $\rho_c = 1.88 \times 10^{-26} h^2\, {\rm kg/m^3}$, $h$ being the dimensionless Hubble parameter. Accounting for a possible entropy non-conservation from evaporation to now, namely $\bar \alpha (s a^3)_{ev}=(s a^3)_0$, we have $Y_{DM}(t_{0})=Y_{DM}(t_{ev})/\bar \alpha$.
Hence
\beq
\Omega_{DM} 
= \frac{m_{DM}  }{\rho_c} n_{DM}(t_{ev})  \frac{s(t_0)}{\bar \alpha \,s(t_{ev})}
= \frac{m_{DM}  }{\rho_c}    n_{DM}(t_{ev})    \left( \frac{a(t_{ev})}{a(t_0)}     \right)^3  \, .
\label{eq-ODM}
\eeq

The DM number density at evaporation can be calculated by integrating over all momenta the DM spectrum at evaporation. Using the definition of the psd given in eq.\,(\ref{eq-defpsd}), we have
\beq
n_{DM}(t_{ev}) 
=  \int d^3(cp) \left. \frac{d n_{DM}}{d^3(cp)}  \right|_{t_{ev}} =  \int d^3(cp) \, g_{DM} f_{DM}(p(t_{ev}),t_{ev})
\eeq
\beq
= (k_B  T_{DM}(t_{ev}) )^3  \int dq  \,4\pi \,q^2  \, g_{DM}\, f_{DM}(q) \nonumber
= (k_B T_{DM}(t_{0}))^3 \left(  \frac{a(t_0)}{a(t_{ev})}  \right)^3  \int dq  \,4\pi q^2  \, g_{DM} f_{DM}(q)\,,
\eeq
where in the second-to-last equality we changed the integration variable to $q$ using eq.\,(\ref{eq-defqBH}) and in the last one we exploited the fact that 
$ T_{DM}(t_{ev})
$ scales as the inverse of the scale factor.

Inserting the last result in eq.\,(\ref{eq-ODM}), the DM abundance is simply given by
\beq \label{eq:omega_psd}
\Omega_{DM} 
= \frac{m_{DM}  }{\rho_c}     
(k_B T_{DM}(t_{0}))^3  \int dq  \,4\pi \,q^2  \, g_{DM} f_{DM}(q)\, ,
\eeq
where $T_{DM}(t_{0})$ is given by eq.\,(\ref{eq-edec}) in units of $T_\gamma(t_0)$.

\subsection{DM from primordial BHs}

Assuming that the DM is fully given by a stable DM candidate from the BH evaporation, we now calculate the required value for its mass. Our analysis is, to our knowledge, the first that accounts for the differences due to the spin of the DM candidate.

For comparison, we exploit both the analytical results for the psd obtained in the geometrical optics limit and the numerical ones obtained using \texttt{BlackHawk}. As for the dof, we consider those of the SM, with the addition of the DM candidate ones. For the DM dof we make a minimal choice, considering just those associated to polarization. For the geometrical optics approximation, we consider a boson B with $g_{DM}=1$, and a fermion F with $g_{DM}=2$. For \texttt{BlackHawk} we consider massive DM particles with $g_{DM} = 1,2,3,4,5$ for $s=0,1/2,1,3/2,2$ respectively.

For $M_{BH}  \lesssim 10^7$ g, we obtain the following results:
for full BH domination
\beq
m_{DM} c^2=  \bar  \alpha \,\frac{\Omega_{DM} }{0.25}  \frac{0.1}{  \left(  \frac{g_{*,S} (t_r)}{ g_{*,S} (t_{ev})} \right) }
 \, \left( \frac{ M_{BH} }{1 \, {\rm g}}   \right)^{1/2} \,\,  \bar m_{BH}c^2 \, ,
  \label{eq-MDM-BH}
\eeq
while for radiation domination 
\beq
m_{DM} c^2=  \bar  \alpha \,\frac{\Omega_{DM} }{0.25}  \frac{0.1}{  \left(  \frac{g_{*,S} (t_r)}{ g_{*,S} (t_{ev})} \right) }
 \, \left( \frac{ M_{BH} }{1 \, {\rm g}}   \right)^{1/2} \,\frac{\bar \beta}{\beta}\,\,  \bar m_{R}c^2 \, ,
 \label{eq-MDM-R}
\eeq 
where the values of $\bar m_{BH,R}$ are collected in table\,\ref{table-MDM}.

\begin{table}[ht]
\centering
\vskip .5 cm
\begin{tabular}{c || c c || c c c c c}
 & B & F & $s=0$  &$s=1/2$  &$s=1$  &$s=3/2$ & $s=2$  \\ [0.5ex] 
\hline\hline
$\bar m_{BH}c^2 / {\rm MeV}$ & 0.114 & 0.076 & 0.112  & 0.155 & 0.344 & 
2.28 & 2.59\\ [1ex]
\hline
$\bar m_{R} c^2/ {\rm MeV}$ & 0.086 & 0.057 & 0.084 & 0.116 &  0.259 &
1.71 & 1.94 \\ [1ex]
\end{tabular}
\caption{Values of $\bar m_{BH}$ and $\bar m_{R}$, calculated: in the geometrical optics approximation, for B and F; from \texttt{BlackHawk}, with increasing values of the spin. Numerical errors are of order of a few percent.}
\label{table-MDM}
\vskip .5 cm
\end{table}

The scaling with the BH mass of eqs.\,(\ref{eq-MDM-BH}) and (\ref{eq-MDM-R}) slightly breaks at $M_{BH} \gtrsim 10^7$ g, where one obtains smaller values for $\bar m_{BH/R}$, as an effect of the decrease of $g_*({t_{ev}})$, see fig.\,\ref{fig-bbeta}. For instance, for $M_{BH}=10^8 (10^9)$ g, the suppression in $\bar m_{BH/R}$ is by a factor of $0.74 (0.56)$.

\begin{figure}[h!]
\vskip .0 cm 
\begin{center}
 \includegraphics[width=9.6cm]{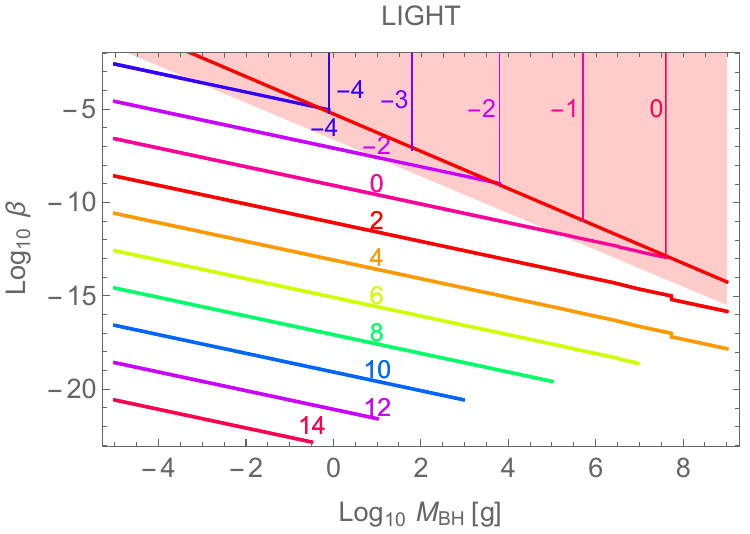}
\end{center}
\caption{\baselineskip=15 pt \small  
Isocountours of ${\rm Log}_{10} \, (m_{DM} c^2[{\rm GeV}])$ for a boson within the geometrical optics approximation, with regions of BH and radiation domination. 
From  \cite{Masina:2020xhk}.  }
\label{fig-DM}
\vskip .2 cm
\end{figure}

As expected, the result for $s=0$ from \texttt{BlackHawk} is consistent with the geometrical optics limit for a boson.
It then perfectly matches with previous literature results obtained with the alternative strategy based on the geometrical optics approximation (see \emph{e.g.} \cite{Masina:2020xhk} and references therein), which we report in fig.\,\ref{fig-DM} (taken form \cite{Masina:2020xhk}) for an easy visualization. The masses $\bar m_{BH}$ and $\bar m_{R}$ are really close in the geometrical optics limit for the bosonic case and for the spin 0 numerical result, the last being slightly lower than the previous. This difference depends on the precise shape of the peak in the psd. For other spins, the suppression of the psd compared to geometrical optics limit causes a clear increase of the masses $\bar m_{BH/R}$ compared to this limit.

Notice that for radiation domination with $\beta=\bar \beta$, the values of $\bar m_{BH/R}$ are systematically smaller by a factor of about $0.75$ with respect to BH domination. Such difference could not be appreciated within the alternative strategy used in the previous literature, as the latter method does not make any difference between the two scenarios. This discrepancy is indeed due to the difference of the ratio of the scale factors in the integrands, not to the slight difference in $T_R(t_{ev})$ which was already included in previous literature.

The strong increase in the masses for increasing values of the spin allows to hope to escape bounds from structure formation.
In the following section we will study in detail the impact on structure formation of these warm DM candidates, assuming entropy conservation, $\bar \alpha=1$. Before doing this, and in view of the comparison to be done in the next section, it is useful to recall the relevant formulas for the DM abundance for a thermal DM candidate.

\subsection{DM from early decoupled thermal relics}

We consider a fermion $X$, with $g_X=2$, as thermal DM candidate.
It is well known that (see \emph{e.g.} \cite{Viel:2005qj}), 
if decoupling happens just before reheating, so that $T_X(t_0)/T_\nu(t_0)=1$, the value of the DM mass for which $\Omega_{X}=0.25$, would be $m_{X} c^2= 11$ eV; for an early decoupling, when $g_{*,S}(t_d)=106.75$, the temperature of the DM candidate now is suppressed, $(T_X(t_0)/T_\nu(t_0))^3 ={g_{*,S} (t_r)}/{ g_{*,S} (t_d)}/\bar \alpha= 0.1/\bar \alpha$, so that its mass gets enhanced to $m_{X} c^2=110$ eV for $\bar \alpha=1$.

We can summarize as follows
\beq
\Omega_X 
= 0.25  \left( \frac{T_X(t_0)}{T_\nu(t_0)} \right)^3  \frac{m_X c^2}{ 11\, {\rm eV}}
= 0.25\,\frac{1}{\bar \alpha} \,\frac{  \left(  \frac{g_{*,S} (t_r)}{ g_{*,S} (t_d)} \right) }{0.1}\,  \frac{m_X c^2}{ 110\, {\rm eV}} \,,
\label{eq-Oth}
\eeq
or, for a direct comparison with the case of DM from evaporating BHs, 
\beq
m_X c^2 = \bar \alpha\, \frac{\Omega_X }{0.25}  \frac{0.1}{  \left(  \frac{g_{*,S} (t_r)}{ g_{*,S} (t_d)} \right) }\,  110\, {\rm eV} \,. 
\eeq

Notice that for $m_X c^2=3 (10)$ keV, $\bar \alpha=1$ and $\Omega_X=0.25$, decoupling would happen when  ${g_{*,S} (t_r)}/{ g_{*,S} (t_d)} =3.6 (1.1) \times 10^{-3}$, namely $g_{*,S}(t_d) \approx 2900 (9800)$, hence much larger than in the SM.

\section{Constraints from structure formation}

Candidates of DM particles are classified according to their velocity dispersion, which defines a free-streaming length. 
On scales smaller than the free-streaming length, fluctuations in the DM density are erased and gravitational clustering is suppressed. 

The velocity dispersion of cold DM (CDM) particles is by definition so small that the corresponding free-streaming length is irrelevant for cosmological structure formation. 
That of hot DM, \emph{e.g.} ordinary light neutrinos, is only one or two orders of magnitude smaller than the speed of light, 
and smoothes out fluctuations in the total matter density even on galaxy cluster scales, which leads to strong bounds on their mass and density. Between these two limits, there exists an intermediate range of DM candidates generically called warm DM (WDM).

The matter power spectrum $P(k)$ is very sensitive to the presence of warm DM particles with large free-streaming lengths.
Due to their free-streaming velocity, warm DM particles slow down the growth of structure and suppress $P(k)$ on scales smaller than their free-streaming scale. 
The effect of the free-streaming on the matter distribution can
be described by a relative “transfer function”
\beq
T(k) = \left(  \frac{P(k)_{\Lambda WDM}}{P(k)_{\Lambda CDM}} \right)^{1/2}\, ,
\eeq
which is the square root of the ratio of the matter power spectrum in the presence of warm DM to that in the presence of purely cold DM, for fixed cosmological parameters.

For the majority of the cosmological models in which the universe contains only warm DM (in addition to the usual baryon, radiation and cosmological constant components), the transfer function can be approximated by the analytical fitting function (see \emph{e.g.}  \cite{Murgia:2017lwo})
\beq \label{eq:transfer_analytical}
T(k) = \left(1+ (\alpha_{B} k)^{\nu} \right)^{\gamma}\,,
\eeq
where $\alpha_{B}$ (labelling the scale of the break), $\nu$ and $\gamma$ are free parameters sensitive to the details of the warm DM candidate. For pure warm DM models, the combined data on the CMB and the Lyman-$\alpha$ forest \cite{Viel:2005qj} 
provide a lower bound on the scale where the transfer function starts to fall. This lower limit was estimated in \cite{Viel:2005qj} to be $\alpha_B \lesssim 0.11 \,{\rm Mpc}/h$ at the 2$\sigma$ confidence level.

A well known case is the one of thermal relics\footnote{There exist however many other warm DM candidates whose origin is rooted in particle physics, \emph{e.g.} the gravitino.}
as warm DM. In such a case eq.\,\eqref{eq:transfer_analytical} simplifies into (see \emph{e.g.} \cite{Bode:2000gq})
\beq
T_{X}(k) = \left(1+ (\alpha_B k)^{2\nu} \right)^{-5/\nu}\,,
\eeq
where $\alpha_{B}$ is a function of the thermal relic mass and temperature, and the index $\nu$ is fixed ($\nu = 1.12$).

The bound on $\alpha_B$ derived in the pioneering analysis of \cite{Viel:2005qj}, translated into a lower bound on the thermal relic mass, gives $m_X c^2 \gtrsim 0.5$ keV. 
A more recent analysis \cite{Irsic:2017ixq} showed that the lower limit is now $m_Xc^2 \gtrsim  3$ keV. 

\begin{figure}[h!]
\vskip .0 cm 
\begin{center}
 \includegraphics[width=7.6cm]{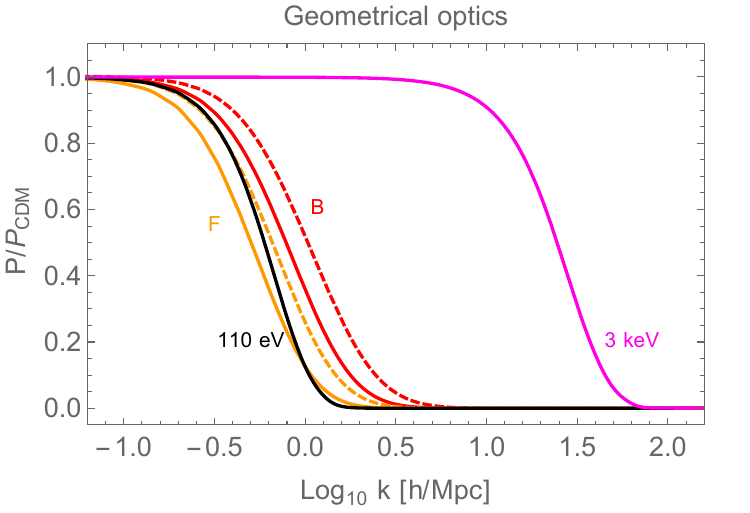}\,\,  \includegraphics[width=7.6cm]{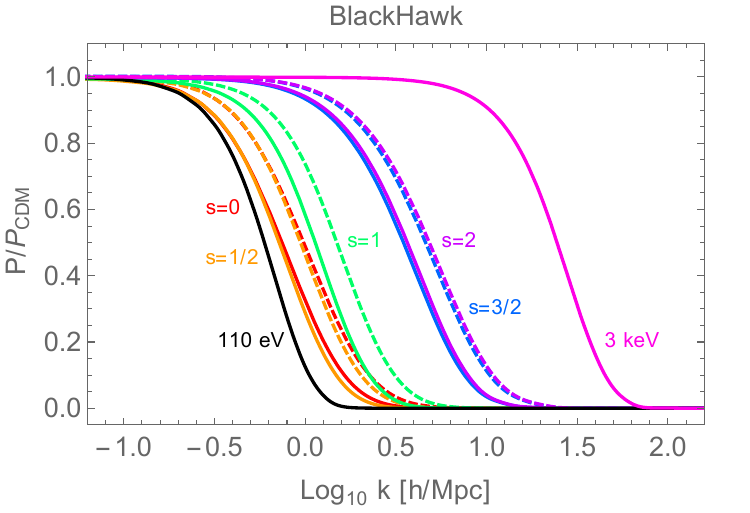} 
  \includegraphics[width=7.6cm]{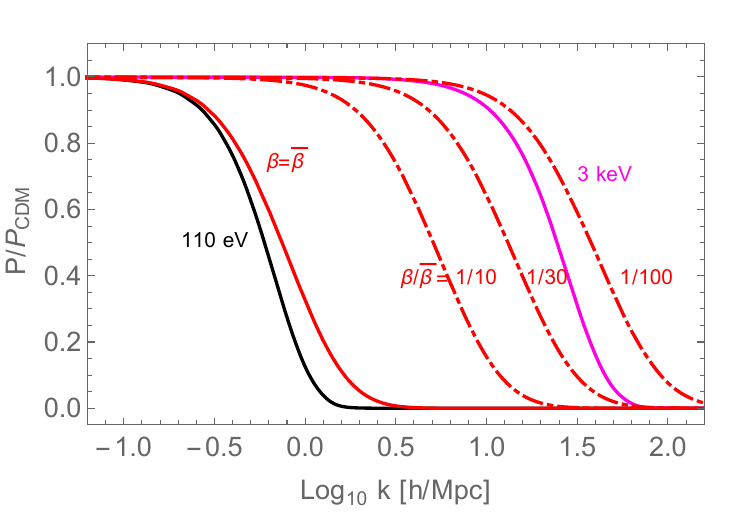} 
\end{center}
\caption{\baselineskip=15 pt \small  
Squared transfer function for various warm DM models providing full DM contribution, computed with \texttt{CLASS}. In all the cases, we have used the same (default) cosmological parameters.
Black: early decoupled thermal fermion ($g_X=2$), with mass $m_X c^2=110$ eV; 
Pink: very early decoupled thermal fermion ($g_X=2$), with mass $m_X c^2=3$ keV. 
Top-Left:  massive boson ($s=0$) and fermion ($s=1/2$) from the evaporation of primordial BHs, for radiation domination with $\beta=\bar \beta$ (solid lines), and for full BH domination (dashed lines), using the psd calculated with the geometrical optics approximation.
Top-Right: the same as Top-Left panel for massive spin $s=0,1/2,1,3/2,2$ particles using the psd calculated with \texttt{BlackHawk}.
Bottom:  $s=0$ warm DM candidate from BH evaporation with radiation domination, for various values of $\beta/\bar\beta$.}
\label{fig-transf}
\vskip .2 cm
\end{figure}

Standard thermal relics (those with mass $m_X c^2 = 11 $ eV) are completely ruled out, as well as early decoupled ones (those with $g_{*}(t_d)=106.75$), with mass $m_X c^2=110$ eV. Only very early decoupled thermal relics could manage to be as heavy as $3$ keV: this would require a huge amount of dof at decoupling. 
We show the transfer function for such early and very early decoupled thermal relics in fig.\,\ref{fig-transf}.

Using the psd's obtained with \texttt{BlackHawk}
(taking the dof of the SM plus those of the DM candidate), for various values of $M_{BH}$ and of the DM particle spin, we calculated with \texttt{CLASS} the associated transfer function, requiring that the DM from the evaporation of primordial BHs accounts for all of the observed DM ($\Omega_{DM}=0.25$). We also require that entropy is conserved, $\bar \alpha =1$. By fixing the BH mass and the DM spin, the mass of the warm DM candidate is univoquely determined, as shown by eqs.\,(\ref{eq-MDM-BH}) and (\ref{eq-MDM-R}).

We show the transfer functions in fig.\,\ref{fig-transf}. The top panels applies to the case of full BH domination (dashed lines) and the case of radiation domination with  $\beta=\bar \beta$ (solid lines): these scenarios give quite similar results. Notice that the transfer functions of fig.\,\ref{fig-transf} actually apply to all values\footnote{The figure of Baldes \emph{et al.} \cite{Baldes:2020nuv} and their eq.\,(5.8) then look at least misleading, as the BH mass, $\Omega_{DM}$ and the DM mass are not independent.} of the BH masses in the range $10^{-5}-10^7$ g, for which the corresponding values of the mass providing $\Omega_{DM}=0.25$ is given by eqs.\,(\ref{eq-MDM-BH}) and (\ref{eq-MDM-R}), according to the spin. It turns out that, even for the higher spins, there is a conflict with the constraints from structure formation (at contrary with the expectations of ref.\,\cite{Lennon:2017tqq}).

For BHs with masses in the range $10^{7}-10^9$ g, the situation is even worse, because the parameter $\bar m_{R/BH}$ is smaller and the ratio $T_{DM}(t_0)/T_\gamma(t_0)$ is larger than in the previous case, being proportional to $1/g_{*,S}(t_{ev})^{1/3}$.

The only possibility left is then radiation domination with a sufficiently small value of $\beta$. In the bottom panel of fig.\,\ref{fig-transf}, we consider radiation domination with increasingly smaller values of $\beta$.
In particular, focusing on the $s=0$ case, we consider the transfer function with $\beta=\bar\beta$ (solid) and compare it with the ones (dot dashed) obtained taking $\beta/\bar\beta = 1/10,1/30,1/100$. We can see that the upper limit on $\beta/\bar\beta$ is about $1/100$: this confirms previous estimates \cite{Fujita:2014hha, Masina:2020xhk} based on a simplified method. The region to be excluded in fig.\,\ref{fig-DM} is thus the same as derived in \cite{Masina:2020xhk}.

For the different values of the spin, $s=0,1/2,1,3/2, 2$, by fitting the right and bottom panels of fig.\,\ref{fig-transf} with eq.\,\eqref{eq:transfer_analytical}, we can derive a general formula for $\alpha_{B}$,
\beq \label{eq:kBpbh}
 \alpha_{B} \simeq \left( \frac{\beta}{\bar \beta}\right)^{0.8} \, (0.95,\, 0.85,\, 0.51,\, 0.14,\, 0.13) \, \frac{\rm Mpc}{h} \,.
\eeq
Since for the transfer function associated to the $3$ keV thermal relic one has $\alpha_{B} \simeq 0.03 \,\frac{\rm Mpc}{h}$, we can derive an upper limit on $\beta / \bar \beta$.
From eq.\,\eqref{eq:kBpbh}, for the different values of the spin, $s=0,1/2,1,3/2,2$, the maximum value of $\beta$ that allows to satisfy the bounds from structure formation turns out to be  
\beq
\frac{\beta}{\bar \beta} \lesssim   \,  (0.013,\, 0.015, \,0.029 ,\, 0.15,\,0.16) \, .
\label{eq-bbb}
\eeq
Our result for the spin 1/2 case agrees with Baldes \emph{et al.} (eq.\,(6.1) of ref.\,\cite{Baldes:2020nuv}), where they obtained $\beta < 0.016 \beta_c$, with their $\beta_c$ being our $\bar \beta$.

\begin{figure}[h!]
\vskip .0 cm 
\begin{center}
 \includegraphics[width=7.6cm]{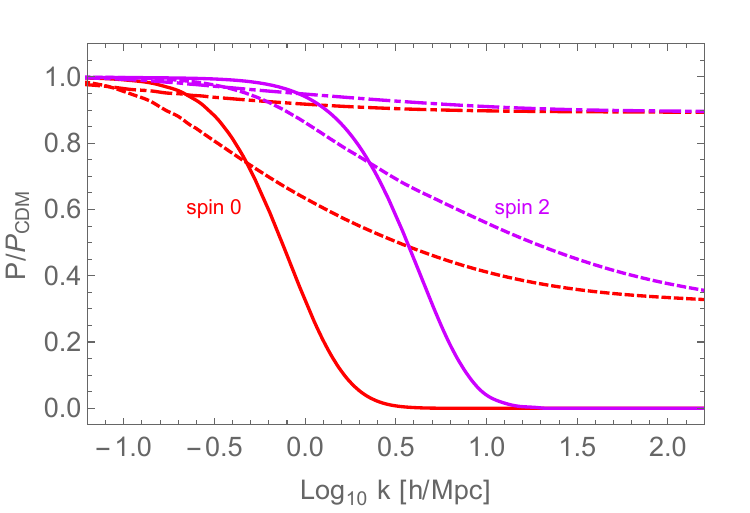}
\end{center}
\caption{\baselineskip=15 pt \small  
Squared transfer function for mixed warm-cold DM models computed with the same cosmological parameters and candidates as fig.\,\ref{fig-transf}, Top-Right panel, for two different spin particles and radiation domination with $\beta = \bar \beta$. Solid lines: $\Omega_{WDM} =  \Omega_{DM, \rm total}$, dashed lines: $\Omega_{WDM} =  1/10 \, \Omega_{DM, \rm total}$, dot-dashed lines: $\Omega_{WDM} =  1/100 \, \Omega_{DM, \rm  total}$.}
\label{fig-transf-mixed}
\vskip .2 cm
\end{figure}

One might also be interested in mixed scenarios with the simultaneous presence of both cold and warm DM, with the warm candidate coming from the evaporation of primordial BHs accounting only partially for the full DM. For the case of thermal relics it has been shown (see \emph{e.g.} \cite{Viel:2005qj}) that in mixed models small-scale structures are not completely erased below $\alpha_{B}$ and the free-streaming effect leads to a step-like transfer function, with the standard $\rm \Lambda CDM$ matter power spectrum recovered in the limit $\Omega_{WDM} \rightarrow 0$. However, in such models the scale of the break $\alpha_{B}$ becomes larger than pure warm DM models, increasing with the inverse of the mass of the DM candidate. We have verified that the same scenario arises with candidates coming from the evaporation of primordial BHs, see \emph{e.g.} fig.\,\ref{fig-transf-mixed}. In such a case the mass of the warm candidate drops as we reduce its abundance, according to eq.\,\eqref{eq:omega_psd}.

\section{Discussion and conclusions}

We improved previous analyses of the constraints on warm DM from primordial BHs evaporation \cite{Fujita:2014hha,Lennon:2017tqq,Masina:2020xhk,Baldes:2020nuv}, by taking into account the effect of the DM spin by means of the code \texttt{BlackHawk} \cite{Arbey:2019mbc}. Assuming that this kind of warm DM provides the full contribution to the observed DM, we calculated the associated transfer function by means of the code \texttt{CLASS} \cite{Lesgourgues:2011re,Blas:2011rf,Lesgourgues:2011rh}.

Contrary to expectations based on \cite{Lennon:2017tqq}, it turns out that, for BH domination, such candidates are excluded for all spin values from $0$ to $2$, in the whole primordial BH mass range $10^{-5}-10^9$ g. Only radiation domination is allowed, if the values $\beta$ are smaller than indicated in eq.\,(\ref{eq-bbb}), according to the warm DM spin.

A couple of possibilities to evade the previous conclusions should be mentioned, which rely on the introduction of additional particles or fields. As suggested in \cite{Fujita:2014hha}, some mechanism providing entropy non conservation and taking place after the evaporation of primordial BHs (like \emph{e.g.} moduli decay) might succeed in saving the warm DM candidates with BH domination. Another possibility to evade the bound would be to have a huge increase in the dof at evaporation. This is not possible to be studied within the present version of the \texttt{BlackHawk} code, but since analytically it is known that $\Omega_X \propto 1/f(M_{BH})^{1/2}$ (see \emph{e.g.} \cite{Masina:2020xhk}), the required increase in the Page function $f(M_{BH})$ would be really huge: a factor of about $10^4$ for $s=0,1/2$.

While the effect of accretion does not help \cite{Masina:2020xhk}, it is plausible that the primordial BHs spin (Kerr case) might help. Primordial BHs are usually thought to form during a radiation dominated era, and thus to have really small spin. However, if formed during a transient matter dominated era, they could have sizeable to near extremal spin \cite{Arbey2020}. The Hawking radiation yield of high spin particles is grealty enhanced by the coupling with the BH spin. Hence, with the same initial density of primordial BHs, a greater fraction of their initial mass could be emitted as high spin warm DM particles.

In this work we considered non-interacting DM from primordial BHs evaporation. Interestingly enough, allowing for self-interacting DM offers the possibility to escape the structure formation bound in the light case for BH domination \cite{Bernal:2020kse}. Thermalization in the DM sector indeed decreases the mean DM kinetic energy and, together with
number-changing processes, can have a strong impact, in particular enhancing the DM relic abundance by several orders of magnitude.

\section*{\Large Acknowledgements}

I.M. acknowledges partial support by the research project TAsP (Theoretical Astroparticle Physics) funded by the Instituto Nazionale di Fisica Nucleare (INFN). \\
G.O. acknowledges support from the Netherlands organization for scientific
research (NWO) VIDI grant (dossier 639.042.730).

\bibliographystyle{elsarticle-num} 
\bibliography{bib} 
\end{document}